\documentclass[pra,amsmath,aps,10pt,superscriptaddress,letterpaper,tightenlines,showpacs]{revtex4}
\usepackage{bm}
\usepackage{stmaryrd}
\usepackage{amsmath}
\usepackage{amssymb}
\usepackage{graphicx}
\usepackage{textcomp}
\usepackage{calrsfs}
\usepackage{yfonts}
\usepackage{color}

\newcommand{\curl}{{\nabla\times}}
\newcommand{\haf}{{\frac{1}{2}}}

\newcommand{\la}{{\langle}}
\newcommand{\ra}{{\rangle}}

\begin{document}
\title{The energy level shifts and the decay rate of an atom in the presence of a conducting wedge }

\author{Zahra Mohammadi}
\affiliation{Department of Physics,
Faculty of Science, University of Isfahan,
Isfahan, Iran}
\author{Fardin Kheirandish}
\email{fkheirandish@yahoo.com}
\affiliation{Department of Physics,
Faculty of Science, University of Isfahan,
Isfahan, Iran}
\begin{abstract}
In the present article explicit expressions for the decay rate and energy level shifts of an atom in the presence of an ideal conducting wedge, two parallel plates and a half-sheet are obtained in the frame work of the canonical quantization approach. The angular and radial dependence of the decay rate for different atomic polarizations of an excited atom and also of the energy level shifts are depicted and discussed. The consistency of the present approach in some limiting cases is investigated by comparing the relevant results obtained here to the previously reported results.
\end{abstract}
\maketitle
\section{Introduction}
An immediate consequence of the quantization of the electromagnetic field is the occurrence of the field fluctuation in the vacuum state. The effect of vacuum fluctuations on atomic systems leads for example to observable phenomena like spontaneous emission or atomic energy level shifts \cite{Vogel,Agarwal}. These radiative properties are explained as the reaction of atom against the existence of the zero-point field.

In quantum field theory the existence of any modification in the presence of an environment is an interesting subject, which have been widely studied. In principle, the  calculations of these radiation properties in the presence of an environment, therefore, become the search for the quantized electromagnetic field in the presence of material fields in order to have a correct picture of fluctuating induced effects on atomic systems.

Therefore we should quantize the electromagnetic field in the presence of material fields \cite{Huttner, Kheirandish1, Kheirandish2, Kheirandish3} in order to have the explicit form of the field operators. A similar situation arise in static or dynamical Casimir effects which is a consequence of constrained vacuum
fluctuations imposed by boundary conditions on macroscopic objects \cite{Milton}. The presence of a boundary surface gives rise to alterations of vacuum field fluctuations and accordingly the energy level shifts and the decay rate of atomic systems change as the atom changes its position with respect to boundary surfaces \cite{Babiker,Barton1,Barton2,Meschede,Philpott,Green,Barton3,Barton4}.
The formalisms using different methods for the only dipole decay rate was explained in other literatures \cite{Rosa, Zhao, Skipsey}. In this paper, we use another approach that describe both of the decay rate and the energy level shifts of a an atom in terms of the imaginary part of the dimensionless vector potential Green's function. We will see that for the dipole decay rate agreement with results of the other approaches are found at the end.

As expected, the de-excitation process for different wedges can also occur for an atom inside a wedge. This phenomenon suggesting that the work presented here is applicable to the area of quantum information processing and the system might serve as a qubit \cite{qinformation}.

In the present work, the decay rate and energy level shifts of an atom in the presence of a perfect conducting wedge, in the frame work of canonical quantization, is investigated \cite{Kheirandish1, Kheirandish2, Kheirandish3}. The formalism applied here is based on the explicit components of the Green's dyadic of the electromagnetic field in the presence of material fields. Various physically important limiting cases of a wedge such as a plane sheet, parallel plates and a half-sheet are considered.

The paper is organized as follows. In Sec. II, the basic material and formulas are presented. In this section, explicit formulas for calculating the decay rate of an initially excited atom and atomic energy level shifts in the presence of a perfectly conducting wedge are given. In Sec. III, The decay rate and energy level shifts of an atom in terms of its polar coordinates $(r_0,\theta_0)$ inside the wedge are depicted and discussed. In Sec. IV, the decay rate of an atom for both parallel and perpendicular polarizations are depicted in terms of the scaled distance of the atom to one plane and also in distance between the plates and the results are discussed. In Sec. V, the decay rate and energy level shifts for an atom in the presence of a perfect conducting half-plane are obtained and discussed. Finally, we discuss the main results and conclude in Sec. VI.
\section{Basic formulae}
\subsection{Effective products for the electric field}
In a general, linear, isotropic magnetodielectric medium, the electric field satisfies the equation \cite{Kheirandish3}
\begin{equation}\label{general}
 \curl(\frac{\mathbf{1}}{\mu}\curl\mathbf{E})-\frac{\omega^2}{c^2}\,\epsilon\mathbf{E} =
\mu_0\omega^2\mathbf{P}^{N}+i\mu_0\omega\curl\mathbf{M}^{N},
\end{equation}
where $\mathbf{P}^{N}(\mathbf{r},\omega)$ and $\mathbf{M}^{N} (\mathbf{r},\omega)$ are polarization and magnetization noise fields and $\epsilon(r,\omega)$ and $\mu(r,\omega)$ are dimensionless permittivity and permeability of the medium, respectively. The constant $\mu_0$ is the permeability of the vacuum. The Green's dyadic $\mathbf{D}(\mathbf{r},\mathbf{r}';\omega)$ of Eq.(\ref{general}) satisfies
\begin{equation}\label{gen-dyad}
 \curl(\frac{\mathbf{1}}{\mu}\curl\mathbf{D})-\frac{\omega^2}{c^2}\,\epsilon\mathbf{D} = \mu_0\omega^2\mathbf{1}\delta(\mathbf{r}-\mathbf{r}').
\end{equation}
Let us assume for simplicity that the medium is nonmagnetic, then from Eq.(\ref{gen-dyad}), the electric field $\mathbf{E}(\mathbf{r},\omega)$ can be written in terms of the polarization noise $\mathbf{P}^N(\mathbf{r},\omega)$ as
\begin{equation}\label{main-E}
 \mathbf{E}(\mathbf{r},\omega)=\int d\mathbf{r}'\,\mathbf{D}(\mathbf{r},\mathbf{r}';\omega)\cdot\mathbf{P}^N(\mathbf{r}',\omega).
\end{equation}
The Green's dyadic in real time can be obtained from the Fourier transform
\begin{equation}
 \mathbf{D}(\mathbf{r},\mathbf{r}';\tau)=\int_{-\infty}^{+\infty}\frac{d\omega}{2\pi}\,e^{-i\omega\tau}\,\mathbf{D}(\mathbf{r},\mathbf{r}';\omega),
\end{equation}
where $\tau=t-t'$. Eq.(\ref{main-E}) in space-time can be written as \cite{milton, DeRaad, casimir p}
\begin{equation}\label{main-E-real}
 \mathbf{E}(\mathbf{r},t)=\int\int d\mathbf{r}'dt'\,\mathbf{D}(\mathbf{r},\mathbf{r}';t-t')\cdot\mathbf{P}^N(\mathbf{r}',t').
\end{equation}
The quantum mechanical monochromatic expectation values are related to the Green's dyadic trough \cite{Philbin}
\begin{eqnarray}\label{effective-pro}
  \la E_i (\mathbf{r}) E_j (\mathbf{r}')\ra_\omega &=& \frac{\hbar\mu_0}{\pi}\omega^2 D_{ij} (\mathbf{r},\mathbf{r}':\omega), \\
  \la H_i (\mathbf{r}) H_j (\mathbf{r}')\ra_\omega &=& \frac{\hbar}{\pi\mu_0}\lim_{\mathbf{r}'\rightarrow \mathbf{r}}
  \frac{1}{\mu^2 (\mathbf{r},\omega)}[\nabla\times D(\mathbf{r},\mathbf{r}';\omega)\times\nabla']_{ij}.
\end{eqnarray}
Here we are considering medium as a perfect conductor so we solve equation (\ref{gen-dyad}) in vacuum with the restriction that the field components on boundaries should satisfy perfect conductor boundary conditions.
\subsection{The decay rate of an initially excited atom}
To find the radiative properties of an atom in the presence of a boundary surface we need the explicit form of the field operators. Finding the explicit form of these operators in a general geometry, due to the intricate structure of field expressions, is a very difficult task or even impossible if we are not invoking to numerical calculations. But one can find an alternative approach to find the radiative properties of an atom without dealing with the explicit forms of the field operators. In this approach we try to find the electromagnetic dyadic tensor satisfying all boundary conditions imposed on the boundaries. Here the boundary which we are interested in is an ideal conducting wedge which include plane and half-plane as limiting cases. Here a very short introduction to the derivation of the basic formulae is given and the details of the calculations can be found for example in \cite{Matloob}.

Up to the dipole approximation, the decay rate of an excited atom is given by Fermi's golden rule
\begin{equation}\label{Fermi}
 \Gamma=\frac{2\pi}{\hbar}\sum_f |\la f|\mathbf{d}_0\cdot\hat{\mathbf{E}}(\mathbf{r}_0,t)|0\ra|^2\,\delta(\omega_f-\omega_0),
\end{equation}
where $\mathbf{r}_0$, $\omega_0$, and $\mathbf{d}_0$ are the position, transition frequency, and dipole moment of the atom, respectively. The kets $|f\ra$ and $|0\ra$ are the final and vacuum states of the electromagnetic field, respectively. If we decompose the electric field to positive and negative frequency parts and make use of the fluctuation-dissipation theorem and Kubo's formula \cite{Kubo}
\begin{equation}\label{Kubo}
 \la 0|\hat{E}^+_\alpha (\mathbf{r},\omega)\,\hat{E}^-_\beta (\mathbf{r}',\omega')|0\ra=
 2\hbar\omega^2\,\mbox{Im} D_{\alpha\beta} (\mathbf{r},\mathbf{r}',\omega)\,\delta(\omega-\omega'),
\end{equation}
one can find the decay rate of an initially excited atom as \cite{Matloob}
\begin{equation}\label{decay}
  \Gamma=\frac{2}{\hbar^{2}}\omega^{2}\mbox{Im}[\mathbf{d}_0\cdot\mathbf{D}(\mathbf{r}_{0},
  \mathbf{r}_{0},\omega)\cdot\mathbf{d}_0],
\end{equation}
where $\mathbf{D}(\mathbf{r}_{0},\mathbf{r}_{0},\omega)$ is the Green's tensor of the electromagnetic field in the presence of boundaries with components $D_{\alpha\beta}$ appearing in (\ref{Kubo}). For dimensional considerations, usually Green tensor is written in terms of the dimensionless Green's tensor $\mathcal{D_{\alpha\beta}}(\mathbf{r},\mathbf{r}',\omega)$, as
\begin{equation}\label{2}
  D_{\alpha\beta}(\mathbf{r},\mathbf{r}',\omega)=\frac{\omega}{4\pi\varepsilon_{0}c^{3}}
  \mathcal{D_{\alpha\beta}}(\mathbf{r},\mathbf{r}',\omega),
\end{equation}
where $\varepsilon_{0}$ and $c$ are the permittivity and the velocity of light in free space respectively. Throughout the paper summation convention is assumed i.e., repeated indices are summed over the three Cartesian coordinates $x,y,z$. In the absence of boundaries or material fields, the decay rate of an excited atom turns out to be
\begin{equation}\label{decay2}
  \Gamma_{0}=\frac{d_0^{2}\omega_{0}^{3}}{3\pi\varepsilon_{0}\hbar c^{3}}.
\end{equation}
By inserting Eqs.(\ref{2},\ref{decay2}) into (\ref{decay}), we finally find
\begin{equation}\label{Gama}
  \Gamma_{\alpha}=\frac{3}{2}\Gamma_{0}\mbox{Im}[\mathcal{D_{\alpha\alpha}}(\mathbf{r}_0,\mathbf{r}_0,\omega_{0})],
\end{equation}
where the subscript $\alpha$ refers to the different orientations of the dipole moment of the atom.
\subsection{Energy level shift}
The presence of the quantum vacuum fluctuations is responsible for fluctuations of the position of an atomic electron around a mean value as
\begin{equation}\label{R}
  \hat{R}(\mathbf{r},t)=\hat{R}_0 (\mathbf{r},t)+\triangle \hat{R}(\mathbf{r},t),
\end{equation}
where $\hat{R}_0 (\mathbf{r},t)$ is the mean value position and $\triangle \hat{R}(\mathbf{r},t)$ is the fluctuating part. Fluctuations of the position cause fluctuations in the potential, using Taylor expansion
\begin{eqnarray}\label{V}
 V(\hat{R}_0+\triangle \hat{R}) &=& V(\hat{R}_0)+(\triangle \hat{R}\cdot\nabla)\,V(\hat{R})\nonumber\\
 && +\haf (\triangle \hat{R}\cdot\nabla)^2\,V(\hat{R})+\cdots,
\end{eqnarray}
where
\begin{equation}\label{Vo}
  V(\hat{R}_0)=-\frac{Z e^2}{4\pi\epsilon_0 R_0},
\end{equation}
is the potential at the mean value position. Now the energy level shift of the atomic energy state $|n\ra$ can be obtained by evaluating the expectation value of the leading term in (\ref{V}) as
\begin{eqnarray}\label{DE}
  \Delta E_n &=& \haf \la (\triangle \hat{R}\cdot\nabla)^2\,V(\hat{R})\ra\nonumber\\
  &=& \frac{Z e^2}{8\pi\epsilon_0}\,Q_{\alpha\beta}\,\la [\triangle \hat{R}(\mathbf{r}_0,t)]_\alpha[\triangle \hat{R}(\mathbf{r}_0,t)]_\beta\ra,
\end{eqnarray}
where
\begin{equation}\label{Q}
  Q_{\alpha\beta}=-\langle n|\frac{\partial^{2}}{\partial x_{\alpha}\partial x_{\beta}}\frac{1}{R}|n\rangle.
\end{equation}
In the dipole approximation, $\triangle \hat{R}(\mathbf{r},t)$ satisfies the Langevin equation
\begin{equation}\label{Langevin}
  m\frac{d^2}{d t^2}\triangle \hat{R}(\mathbf{r},t)+m\Gamma \frac{d}{d t}\triangle \hat{R}(\mathbf{r},t)=-e\hat{E}(\mathbf{r},t),
\end{equation}
where $m$ is the mass of the electron and the parameter $\Gamma$ is the damping constant defined by the Bethe's average excitation energy
\begin{equation}\label{Bethe}
  \Gamma=c\gamma=|E_n-E_m|_{\mbox{av}}/\hbar=17.8 R_\infty/\hbar,
\end{equation}
and $R_\infty$ is the Rydberg unit of energy. Eq.(\ref{Langevin}) can be solved using Fourier transform, inserting the solution into Eq.(\ref{DE}) and using Eq.(\ref{Kubo}), we finally find
\begin{equation}\label{En}
 \Delta E_{n}=\frac{ze^{4}\hbar}{32\pi^{3}\varepsilon_{0}^{2}c^{3}m^{2}}\int_{0}^{\frac{mc}{\hbar}}\frac{q}{q^{2}+
  \gamma^{2}}Q_{\alpha\beta}\mbox{Im}[\mathcal{D_{\alpha\beta}}(\mathbf{r_{0}},\mathbf{r_{0}},\omega)]dq,
\end{equation}
where $q=\omega/c$. The cutoff frequency $mc/\hbar$ in Eq.(\ref{En}) which is the Compton wavelength of the electron, is needed due to the validity of the dipole approximation applied in Eq.(\ref{Langevin}) \cite{Matloob}. The main ingredient of the basic formulae (\ref{decay2}) and (\ref{En}) is the dimensionless Green tensor. In the next section the decay rate of an initially excited atom and also the energy level shift are obtained for an atom located at an arbitrary point inside a perfectly conducting wedge.
\section{The Perfectly Conducting Wedge}
\subsection{The decay rate}
Consider an initially excited atom located at an arbitrary point $P$ inside an infinite wedge with perfect conducting walls and the apex angle $\alpha$, Fig.1. Due to the symmetry along the $z$-component, the point $P$ is determined by $(r_0,\theta_0)$ in polar coordinates system. The Green tensor in the presence of a wedge with perfect conducting walls is now a text book problem and the interested reader can find the details of its derivation for example in \cite{Brevik, Tai1, Tai2}. According to Eq.(\ref{Gama}), the relevant components are the diagonal components $\mathcal{D}_{rr}, \mathcal{D}_{\theta\theta}$, and $\mathcal{D}_{zz}$  given by \cite{Brevik, Tai1, Tai2}
\begin{eqnarray}\label{20}
  \mathcal{D}_{rr} &=&\frac{-2ip}{ q^3}\int\limits_{-\infty}^{+\infty} dk \,e^{ik(z-z')}\,\acute{\sum_{m=0}^{\infty}}\,\bigg[\frac{q^2m^2p^2}{\eta ^2\,rr'}J_{mp}(\eta r)\,H_{mp}^{(1)}(\eta r')+k^2\, J_{mp}'(\eta r)\,H_{mp}'^{(1)}(\eta r')\bigg]\sin(mp\theta)\sin(mp\theta'),\nonumber\\
  &=&\frac{-2ip}{ q^3}\bigg[\frac{1}{rr'}\frac{\partial}{\partial \theta}\frac{\partial}{\partial \theta'}\int_{-\infty}^{+\infty}\frac{q^2 dk}{\eta ^2} \,e^{ik(z-z')}\,\acute{\sum_{m=0}^{\infty}}\,J_{mp}(\eta r)\,H_{mp}^{(1)}(\eta r')\cos(mp\theta) \cos(mp\theta')\nonumber\\
  &&+ \frac{\partial}{\partial r}\frac{\partial}{\partial r'}\int\limits_{-\infty}^{+\infty}\frac{k^2 dk}{\eta ^2} \,e^{ik(z-z')}\,\acute{\sum_{m=0}^{\infty}}\,J_{mp}(\eta r)\,H_{mp}^{(1)}(\eta r')\sin(mp\theta)\sin(mp\theta')\bigg],\nonumber\\
  &=&\frac{-ip}{ q^3}\bigg[\frac{1}{rr'}\frac{\partial}{\partial \theta}\frac{\partial}{\partial \theta'}\int_{-\infty}^{+\infty}\frac{q^2 dk}{\eta ^2} \,e^{ik(z-z')}\,\acute{\sum_{m=0}^{\infty}}\,J_{mp}(\eta r)\,H_{mp}^{(1)}(\eta r')\,[\cos(mp(\theta+\theta'))+\cos(mp(\theta-\theta')]\nonumber\\
  &&+ \frac{\partial}{\partial r}\frac{\partial}{\partial r'}\int\limits_{-\infty}^{+\infty}\frac{k^2 dk}{\eta ^2} \,e^{ik(z-z')}\,\acute{\sum_{m=0}^{\infty}}\,J_{mp}(\eta r)\,H_{mp}^{(1)}(\eta r')\,[\cos(mp(\theta-\theta'))-\cos(mp(\theta+\theta')]\bigg],\nonumber\\
\end{eqnarray}
\begin{eqnarray}\label{20}
    \mathcal{D}_{\theta\theta} &=&\frac{-2ip}{q^3}\int\limits_{-\infty}^{+\infty} dk \,e^{ik(z-z')}\,\acute{\sum_{m=0}^{\infty}}\,\bigg[\frac{k^2m^2p^2}{\eta ^2\,rr'}J_{mp}(\eta r)\,H_{mp}^{(1)}(\eta r')+q^2\, J_{mp}'(\eta r)\,H_{mp}'^{(1)}(\eta r')\bigg]\cos(mp\theta)\,\cos(mp\theta'),\nonumber\\
  &=&\frac{-2ip}{q^3}\bigg[\frac{1}{rr'}\frac{\partial}{\partial \theta}\frac{\partial}{\partial \theta'}\int_{-\infty}^{+\infty}\frac{k^2 dk}{\eta ^2}\,e^{ik(z-z')}\,\acute{\sum_{m=0}^{\infty}}\,J_{mp}(\eta r)\,H_{mp}^{(1)}(\eta r')\sin(mp\theta) \sin(mp\theta')\nonumber\\
  &&+ \frac{\partial}{\partial r}\frac{\partial}{\partial r'}\int\limits_{-\infty}^{+\infty}\frac{q^2 dk}{\eta ^2} \,e^{ik(z-z')}\,\acute{\sum_{m=0}^{\infty}}\,J_{mp}(\eta r)\,H_{mp}^{(1)}(\eta r')\cos(mp\theta)\cos(mp\theta')\bigg],\nonumber\\
  &=&\frac{-ip}{q^3}\bigg[\frac{1}{rr'}\frac{\partial}{\partial \theta}\frac{\partial}{\partial \theta'}\int_{-\infty}^{+\infty}\frac{k^2 dk}{\eta ^2} \,e^{ik(z-z')}\, \acute{\sum_{m=0}^{\infty}}\,J_{mp}(\eta r)\,H_{mp}^{(1)}(\eta r')\,[\cos(mp(\theta-\theta'))-\cos(mp(\theta+\theta')]\nonumber\\
  &&+ \frac{\partial}{\partial r}\frac{\partial}{\partial r'}\int\limits_{-\infty}^{+\infty}\frac{q^2 dk}{\eta ^2} \,e^{ik(z-z')}\,\acute{\sum_{m=0}^{\infty}}\,J_{mp}(\eta r)\,H_{mp}^{(1)}(\eta r')\,[\cos(mp(\theta+\theta'))+\cos(mp(\theta-\theta')]\bigg],\nonumber\\
\end{eqnarray}
\begin{eqnarray}\label{20}
  \mathcal{D}_{zz} &=&\frac{-2ip}{q^3}\int\limits_{-\infty}^{+\infty} dk\,\eta ^2\,e^{ik(z-z')}\, \acute{\sum_{m=0}^{\infty}}\,J_{mp}(\eta r)\,H_{mp}^{(1)}(\eta r')\,\sin(mp\theta)\,\sin(mp\theta'),\nonumber\\
  &=&\frac{-ip}{q^3}\int\limits_{-\infty}^{+\infty} dk\, \eta ^2\,e^{ik(z-z')}\,\acute{\sum_{m=0}^{\infty}}\,J_{mp}(\eta r)\,H_{mp}^{(1)}(\eta r')\,[\cos(mp(\theta-\theta'))-\cos(mp(\theta+\theta')],\nonumber\\
\end{eqnarray}
\begin{figure}
    \includegraphics[scale=0.4]{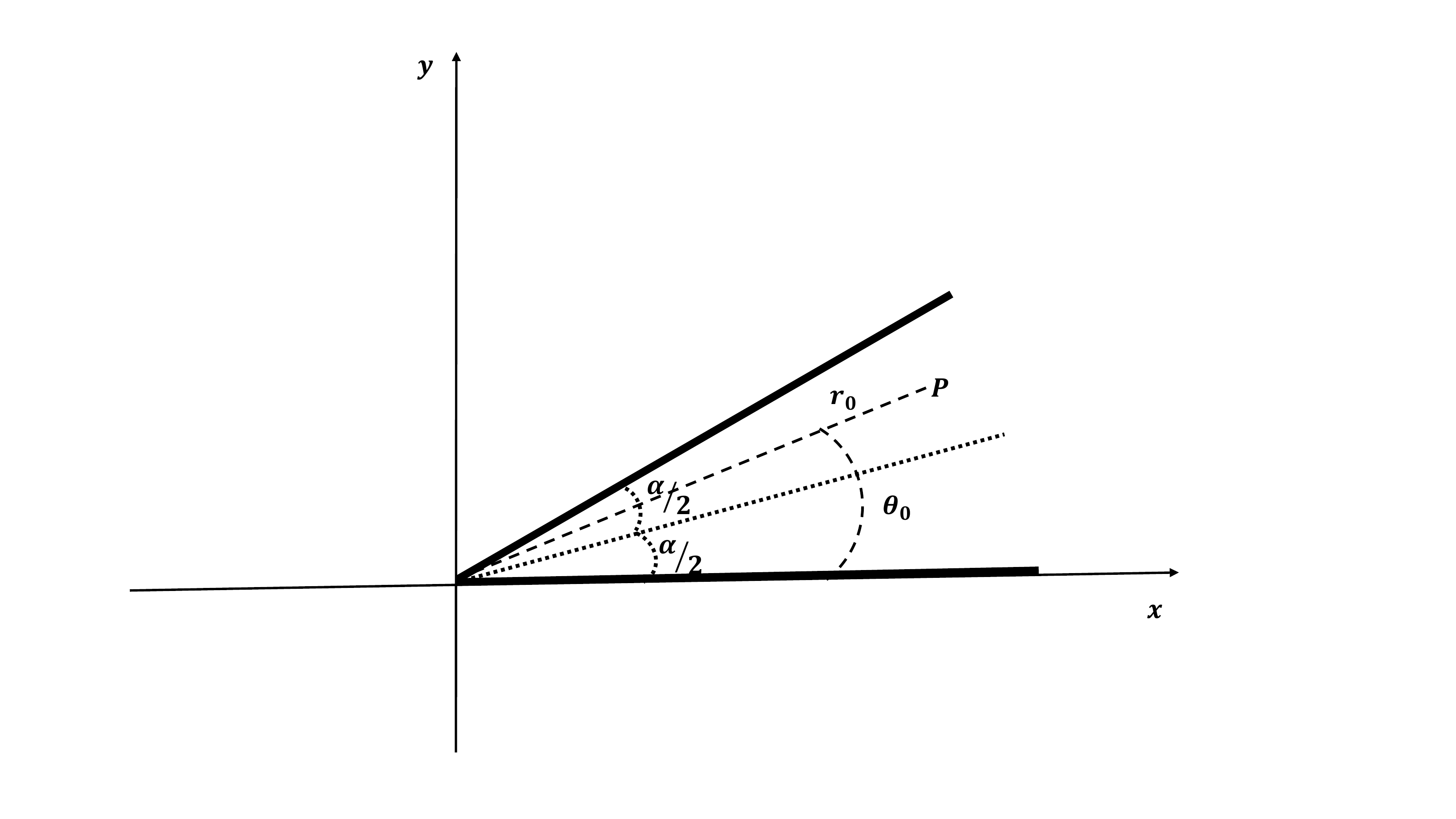}\\
  \caption{(Color online) Atom is located at point P inside a wedge with perfectly conducting walls.}\label{Fig1}
\end{figure}
where $p=\pi/\alpha$ is defined for simplicity and $\eta=\sqrt{q^2-k^2}$. We can henceforth put $z=z'$.  For the case where $\alpha=\pi/n$ where $n$ is a natural number, the parameter $p$ is an integer $(p=n)$, and the summation over $m$ can be done using Graf's addition theorem \cite{gradshteyn}
\begin{equation}\label{Graf}
\sum_{n=0}^{p-1}\,K_{0}(\zeta R_{n})=2p \acute{\sum_{m=0}^{\infty}}\,I_{mp}(\zeta r_{1})\,K_{mp}(\zeta r_{2})\,\cos(mp\phi),
\end{equation}
where
\begin{equation}
 R_{n}=\sqrt{[r_{1}^2+r_{2}^2-2r_{1}r_{2}\,\cos(\phi+2n\pi/p)]}.
\end{equation}
Therefore, using Eq.(\ref{Graf}) and changing the integration variable $u=k/q$, the diagonal components can be written as
\begin{eqnarray}\label{Drr}
  \mathcal{D}_{rr} &=&\frac{2}{\pi q^2}\bigg[\frac{1}{rr'}\,\sum_{n=0}^{p-1} \,\int_{0}^{+\infty}\,\frac{du}{(u^2-1)}\,\frac{\partial}{\partial \theta}\frac{\partial}{\partial \theta'}\,\bigg(K_{0}(\sqrt{u^2-1}\,q R_{1})+K_{0}(\sqrt{u^2-1}\,q R_{2})\bigg)\nonumber\\
  && + \sum_{n=0}^{p-1}\int_{0}^{+\infty} \,\frac{u^2\,du}{(u^2-1)}\,\frac{\partial}{\partial r}\frac{\partial}{\partial r'}\,\bigg(K_{0}(\sqrt{u^2-1}\,q R_{2})-K_{0}(\sqrt{u^2-1}\,q R_{1})\bigg)\bigg],\nonumber\\
\end{eqnarray}
\begin{eqnarray}\label{dtt}
  \mathcal{D}_{\theta\theta}  &=&\frac{2}{\pi q^2}\bigg[\frac{1}{rr'}\,\sum_{n=0}^{p-1} \,\int_{0}^{+\infty} \,\frac{u^2\,du}{(u^2-1)}\,\frac{\partial}{\partial \theta}\frac{\partial}{\partial \theta'}\,\bigg(K_{0}(\sqrt{u^2-1}\,q R_{2})-K_{0}(\sqrt{u^2-1}\,q R_{1})\bigg)\nonumber\\
  && + \sum_{n=0}^{p-1}\, \int_{0}^{+\infty} \,\frac{du}{(u^2-1)}\,\frac{\partial}{\partial r}\frac{\partial}{\partial r'}\,\bigg(K_{0}(\sqrt{u^2-1}\,q R_{1})+K_{0}(\sqrt{u^2-1}\,q R_{2})\bigg)\bigg],\nonumber\\
\end{eqnarray}
\begin{eqnarray}\label{Dzz}
 \mathcal{D}_{zz} &=&\frac{2}{\pi}\,\sum_{n=0}^{p-1}\int_{0}^{+\infty} du\,(u^2-1)\,\bigg(K_{0}(\sqrt{u^2-1}\,q R_{2})-K_{0}(\sqrt{u^2-1}\,q R_{1})\bigg),\nonumber\\
\end{eqnarray}
where
\begin{eqnarray}
R_{1}&=&\sqrt{r^2+r'^2-2rr'\,\cos(\theta+\theta'+2n\pi/p)},\nonumber\\
R_{2}&=&\sqrt{r^2+r'^2-2rr'\,\cos(\theta-\theta'+2n\pi/p)}.
\end{eqnarray}
\begin{figure}
    \includegraphics[scale=0.6]{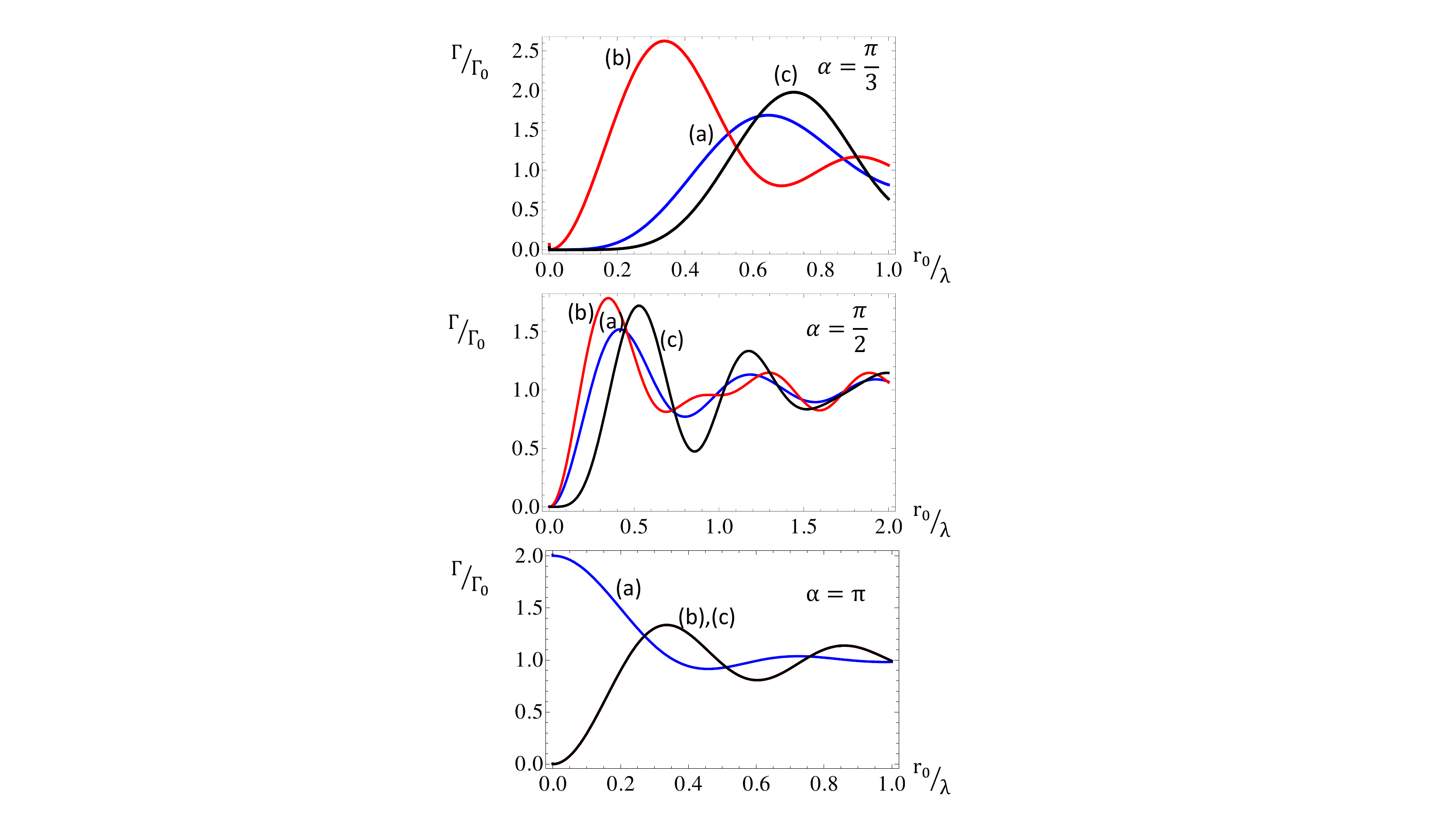}\\
  \caption{(Color online) The dimensionless decay rate of an excited atom for three orientations $(a)\,\frac{\Gamma_{r}}{\Gamma_{0}},\,(b)\,\frac{\Gamma_{\theta}}{\Gamma_{0}},\,(c)\,\frac{\Gamma_{z}}{\Gamma_{0}}$ in terms of dimensionless distance $\frac{r_{0}}{\lambda}$ from the origin along the symmetry line for the wedges with different $\alpha$.}\label{Fig2}
\end{figure}
Now using the formula \cite{gradshteyn}
\begin{equation}\label{gradshteyn}
\int_{0}^{+\infty} dx\,\frac{K_{\nu}(\beta\sqrt{x^2+z^2})}{\sqrt{(x^2+z^2)^\nu}}\,x^{2\mu+1}=\frac{2^\mu\,\Gamma(\mu+1)}{\beta^{\mu+1}\,z^{\nu-\mu-1}}\,K_{\nu-\mu-1}(\beta z),
\end{equation}
and doing some straightforward calculations, we finally find the imaginary part of the diagonal components of Green tensor as
\begin{eqnarray}\label{IDrr}
\mbox{Im}[\mathcal{D}_{rr}(\mathbf{r}_{0},\mathbf{r}_{0},\omega)]&=&\,\sum_{n=0}^{p-1}
\bigg[\big(\frac{\cos x}{x^2}+\frac{\sin x}{x}-\frac{\sin x}{x^3}\big)+\sin^2(\frac{n\pi}{p})\,\big(\frac{\cos x}{x^2}-\frac{\sin x}{x}-\frac{\sin x}{x^3}\big)\nonumber\\&-&\big(\frac{\cos x_{\theta}}{x_{\theta}^2}+\frac{\sin x_{\theta}}{x_{\theta}}-
\frac{\sin x_{\theta}}{x_{\theta}^3}\big)
-\sin^2(\theta+\frac{n\pi}{p})\,\big(\frac{\cos x_{\theta}}{x_{\theta}^2}-\frac{\sin x_{\theta}}{x_{\theta}}-\frac{\sin x_{\theta}}{x_{\theta}^3}\big)\bigg],
\end{eqnarray}
\begin{eqnarray}\label{IDtt}
\mbox{Im}[\mathcal{D}_{\theta\theta}(\mathbf{r}_{0},\mathbf{r}_{0},\omega)]&=&-\,\sum_{n=0}^{p-1}
\bigg[2\,\big(\frac{\cos x}{x^2}-\frac{\sin x}{x^3}\big)-\sin^2(\frac{n\pi}{p})\,\big(\frac{\cos x}{x^2}-\frac{\sin x}{x}-\frac{\sin x}{x^3}\big)\nonumber\\&+&2\,\big(\frac{\cos x_{\theta}}{x_{\theta}^2}-\frac{\sin x_{\theta}}{x_{\theta}^3}\big)
-\sin^2(\theta+\frac{n\pi}{p})\,\big(\frac{\cos x_{\theta}}{x_{\theta}^2}-\frac{\sin x_{\theta}}{x_{\theta}}-\frac{\sin x_{\theta}}{x_{\theta}^3}\big)\bigg],
\end{eqnarray}
\begin{eqnarray}\label{IDzz}
\mbox{Im}[\mathcal{D}_{zz}(\mathbf{r}_{0},\mathbf{r}_{0},\omega)]=\,\sum_{n=0}^{p-1}
\bigg[\big(\frac{\sin x}{x}+\frac{\cos x}{x^2}-\frac{\sin x}{x^3}\big)-\big(\frac{\sin x_{\theta}}{x_{\theta}}+\frac{\cos x_{\theta}}{x_{\theta}^2}-
\frac{\sin x_{\theta}}{x_{\theta}^3}\big)\bigg].
\end{eqnarray}
\begin{figure}
    \includegraphics[scale=0.5]{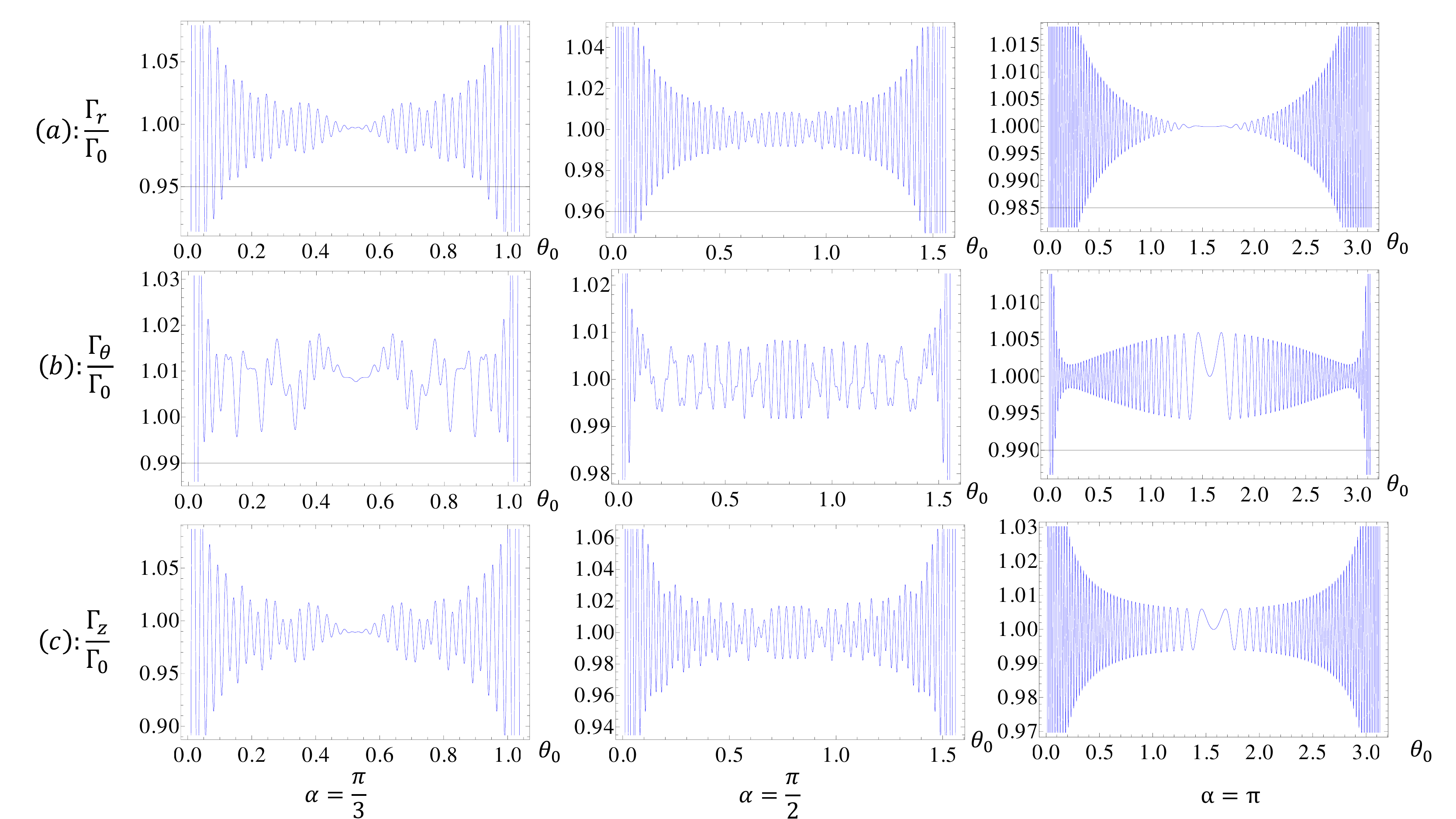}\\
  \caption{(Color online) The dimensionless decay rate for three orientations of the polarization of the excited
atom  $(a)\,\frac{\Gamma_{r}}{\Gamma_{0}},\,(b)\,\frac{\Gamma_{\theta}}{\Gamma_{0}},\,(c)\,\frac{\Gamma_{z}}{\Gamma_{0}}$ in terms of the angle $\theta_0$, and fixed distance $\frac{r_{0}}{\lambda}=20$, for $\alpha=\pi/3,\,\pi/2,\,\pi$.}\label{Fig3}
\end{figure}
By inserting Eqs.(\ref{IDrr},\ref{IDtt},\ref{IDzz}) into (\ref{Gama}), we find
\begin{eqnarray}
\frac{\Gamma_{r}}{\Gamma_{0}}&=&\frac{3}{2}\,\sum_{n=0}^{p-1}
\bigg[\big(\frac{\cos x}{x^2}+\frac{\sin x}{x}-\frac{\sin x}{x^3}\big)+\sin^2(\frac{n\pi}{p})\,\big(\frac{\cos x}{x^2}-\frac{\sin x}{x}-\frac{\sin x}{x^3}\big)\nonumber\\&-&\big(\frac{\cos x_{\theta}}{x_{\theta}^2}+\frac{\sin x_{\theta}}{x_{\theta}}-
\frac{\sin x_{\theta}}{x_{\theta}^3}\big)
-\sin^2(\theta+\frac{n\pi}{p})\,\big(\frac{\cos x_{\theta}}{x_{\theta}^2}-\frac{\sin x_{\theta}}{x_{\theta}}-\frac{\sin x_{\theta}}{x_{\theta}^3}\big)\bigg],
\end{eqnarray}
\begin{eqnarray}\label{gammatheta}
\frac{\Gamma_{\theta}}{\Gamma_{0}}&=&-\frac{3}{2}\,\sum_{n=0}^{p-1}
\bigg[2\,\big(\frac{\cos x}{x^2}-\frac{\sin x}{x^3}\big)-\sin^2(\frac{n\pi}{p})\,\big(\frac{\cos x}{x^2}-\frac{\sin x}{x}-\frac{\sin x}{x^3}\big)\nonumber\\&+&2\,\big(\frac{\cos x_{\theta}}{x_{\theta}^2}-\frac{\sin x_{\theta}}{x_{\theta}^3}\big)
-\sin^2(\theta+\frac{n\pi}{p})\,\big(\frac{\cos x_{\theta}}{x_{\theta}^2}-\frac{\sin x_{\theta}}{x_{\theta}}-\frac{\sin x_{\theta}}{x_{\theta}^3}\big)\bigg],
\end{eqnarray}
\begin{eqnarray}\label{gammaz}
\frac{\Gamma_{z}}{\Gamma_{0}}=\frac{3}{2}\,\sum_{n=0}^{p-1}
\bigg[\big(\frac{\sin x}{x}+\frac{\cos\,x}{x^2}-\frac{\sin x}{x^3}\big)-\big(\frac{\sin x_{\theta}}{x_{\theta}}+
\frac{\cos x_{\theta}}{x_{\theta}^2}-\frac{\sin x_{\theta}}{x_{\theta}^3}\big)\bigg],
\end{eqnarray}
where $x_{\theta}=2r_{0}q\sin(\theta_{0}+n\pi/p)$ and $x=2r_{0}q\sin(n\pi/p)$.

In Fig. 2, the decay rate of an initially excited atom for different polarizations and apex angles is depicted in terms of the distance from the $z$-axis for $\theta_0=\alpha/2$. The decay rates are normalized to the decay rates in free space $\Gamma_0$. Distances are also normalized to wavelength $\lambda$. In all of these diagrams when $r_{0}\gg\lambda$, i.e the atom is far away from the axis, the decay rate tends to the free-space decay rate as expected.

\begin{figure}
    \includegraphics[scale=0.5]{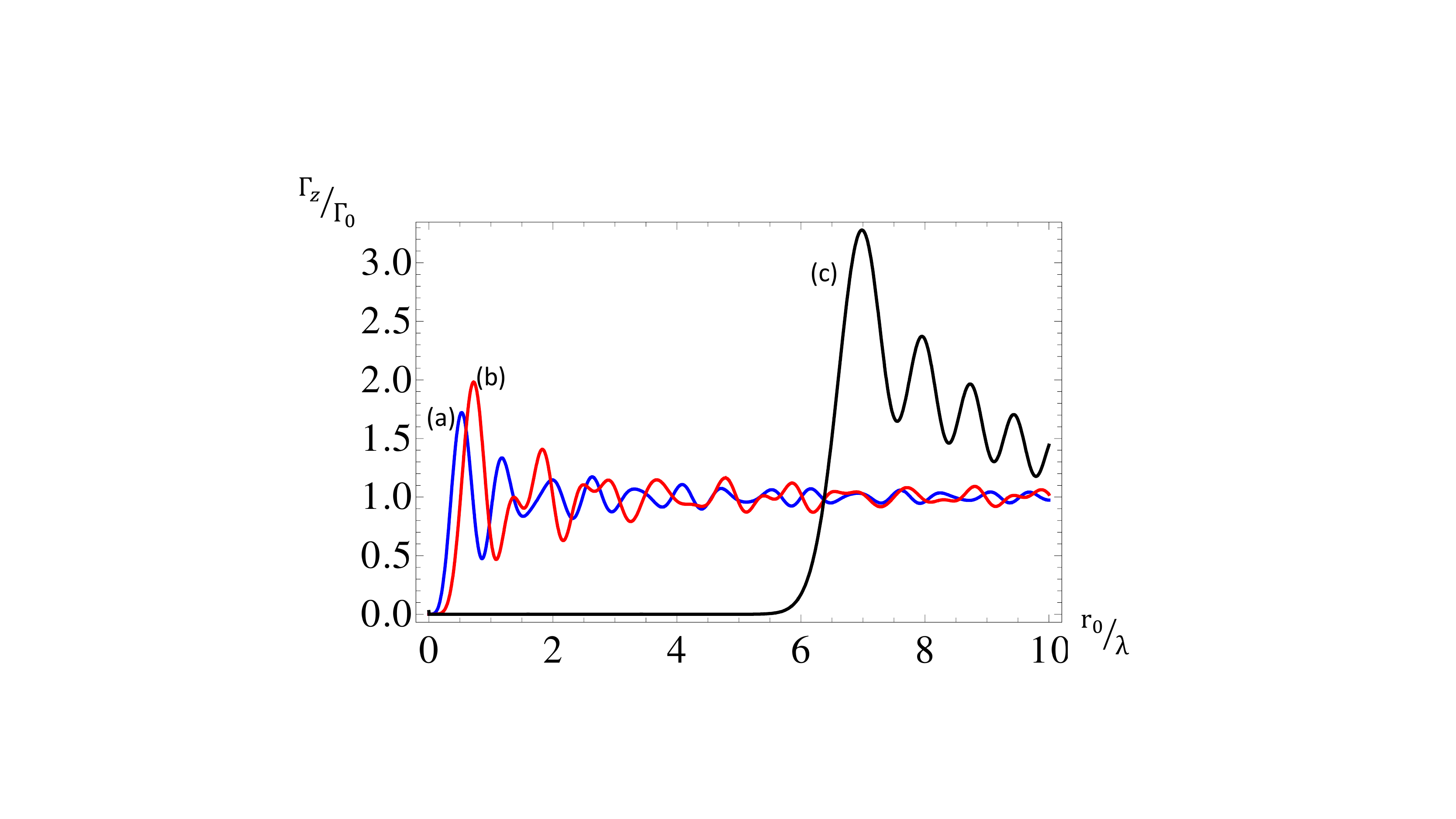}\\
  \caption{(Color online) The dimensionless decay rate $\frac{\Gamma_{z}}{\Gamma_{0}}$ of the excited
atom in terms of dimensionless distance $\frac{r_{0}}{\lambda}$ from the origin along the symmetry line for the wedges with different $\alpha$:
(a)$\,\alpha=\pi/2$, (b)$\,\alpha=\pi/3$ and (c)$\,\alpha=\pi/40$.}\label{Fig4}
\end{figure}

In Fig. 3, the typical behavior of the decay rate in terms of the angle $\theta$ for fixed distance ($r_0=20\lambda$) from the $z$-axis is depicted for different apex angles $\alpha$. As expected, these curves are symmetric around the middle point $\theta_0=\alpha/2$.

In Fig. 4, the damping rate for $z$-polarization is depicted in terms of the scaled distance ($\frac{r_0}{\lambda}$) for different wedge angles. It is seen that for distances from the cusp smaller than a certain value determined by the opening angle of the wedge, there are regions for which the atom will not decay at all and these suppressions are followed by a sudden jump. Also, it should be noted that we are considering perfect conductor here, for a good conductor the influence depth ($\delta$) of the electromagnetic fields inside the conductor is not zero so we can not close to conductors such that the distance to the conductors is comparable to $\delta$. In a real investigation one should consider real metals.

It is interesting to notice that if we consider a wedge with a certain $\alpha$, for example $\alpha=\frac{\pi}{3}$, see Figs.2 and 3, a change in the dipole orientation changes the decay rate from super-radiant to sub-radiant and vice versa, depending on the location of the atomic dipole.

For the special case $\alpha=\pi$, the wedge degenerates into a plane sheet. In this case we find for the decay rates
\begin{eqnarray}
\frac{\Gamma_{\|}}{\Gamma_{0}}&=&\frac{\Gamma_{z}+\Gamma_{\theta}}{\Gamma_{0}},\nonumber\\
&=&1-\frac{3}{2}\bigg[\frac{\sin\,(2\,r_{0}q)}{(2\,r_{0}q)}+\frac{\cos\,(2\,r_{0}q)}{(2\,r_{0}q)^2}-
\frac{\sin\,(2\,r_{0}q)}{(2\,r_{0}q)^3}\bigg],
\end{eqnarray}
and
\begin{eqnarray}
\frac{\Gamma_{\perp}}{\Gamma_{0}}&=&\frac{\Gamma_{r}}{\Gamma_{0}},\nonumber\\
&=& 1-3\bigg[\frac{\cos\,(2\,r_{0}q)}{(2\,r_{0}q)^2}-\frac{\sin\,(2\,r_{0}q)}{(2\,r_{0}q)^3}\bigg],
\end{eqnarray}
which are in agreement with the results reported in \cite{Matloob}.
\subsection{The energy level shift}
\begin{figure}
    \includegraphics[scale=0.5]{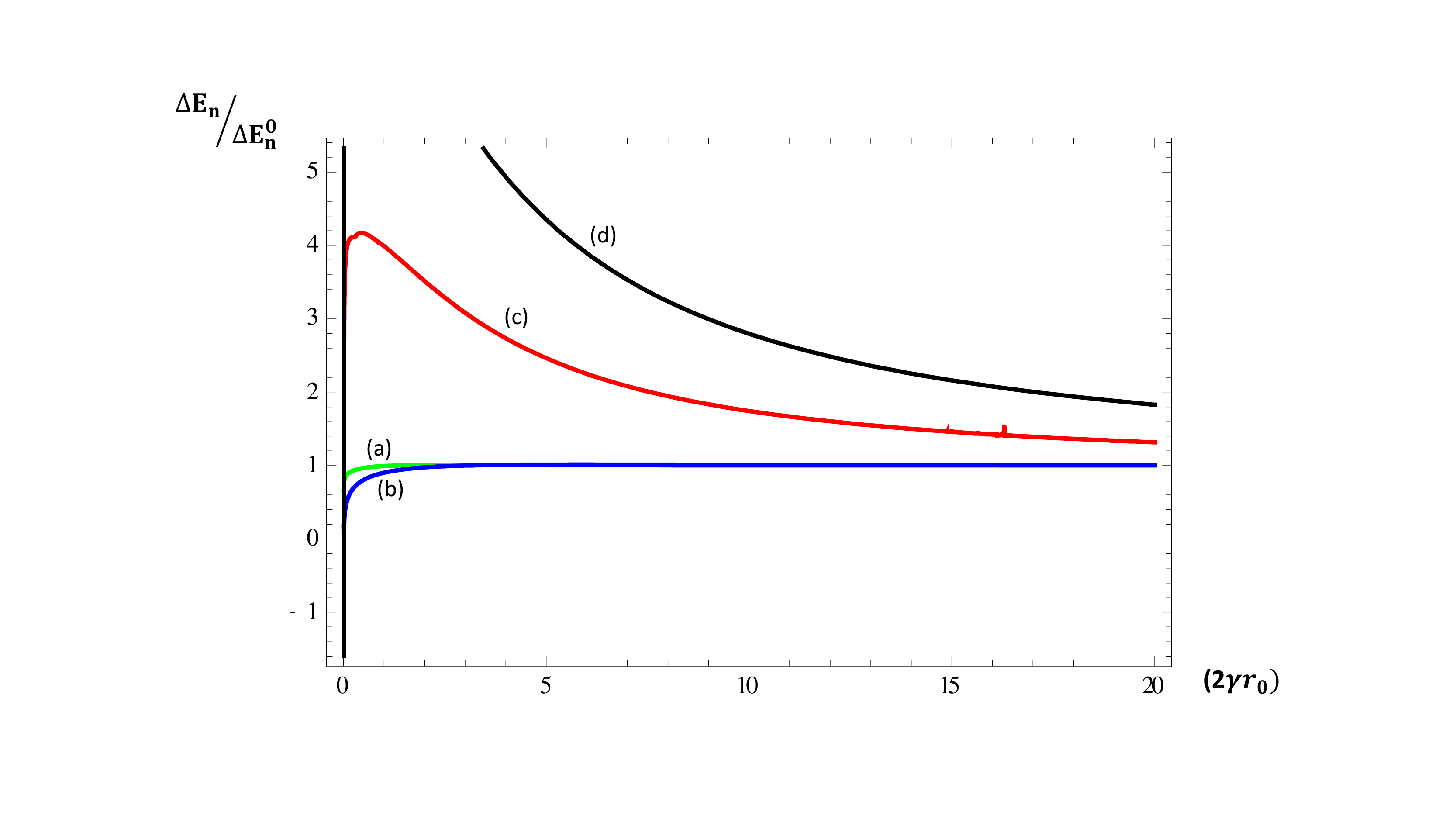}\\
  \caption{(Color online) The relative energy level shifts of an atom inside the wedge in terms of the dimensionless distance ($2\gamma r_0$) along the symmetry line of the wedge with different apex angles $\alpha$: (a) $\alpha=\pi$, (b) $\alpha=\pi/3$, (c) $\alpha=\pi/50$ and (d) $\alpha=\pi/100$.}\label{Fig5}
\end{figure}
In this section we find the energy level shift of an atom placed in an arbitrary point defined by $(r_0,\theta_0)$ in Fig.1. In order to use Eq.(\ref{En}), we note that the repeated indices are summed over the three Cartesian coordinates $\alpha,\beta=x,y,z$. In Cartesian coordinates the Green's dyadic can be obtained from the one in cylindrical coordinates. For the off-diagonal components,as we expected, due to the symmetry of the problem we find
\begin{eqnarray}\label{off1}
\mbox{Im}[\mathcal{D}_{\alpha z}(\mathbf{r}_{0},\mathbf{r}_{0},\omega)]=\mbox{Im}[\mathcal{D}_{z \alpha }(\mathbf{r}_{0},\mathbf{r}_{0},\omega)]=0,
\end{eqnarray}
in which $\alpha=x,y$ and for two other off-diagonal components, it is easy to show that
\begin{eqnarray}\label{off2}
\mbox{Im}[\mathcal{D}_{xy}(\mathbf{r}_{0},\mathbf{r}_{0},\omega)+\mathcal{D}_{yx}(\mathbf{r}_{0},\mathbf{r}_{0},\omega)]\ll
\mbox{Im}[Tr[\mathcal{D}(\mathbf{r}_{0},\mathbf{r}_{0},\omega)]],
\end{eqnarray}
where
\begin{eqnarray}\label{TrD}
Tr[\mathcal{D}]&=&\mathcal{D}_{xx}+\mathcal{D}_{yy}+\mathcal{D}_{zz},\nonumber\\
&=&\mathcal{D}_{rr}+\mathcal{D}_{\theta\theta}+\mathcal{D}_{zz}.
\end{eqnarray}
Therefore, we can rewrite Eq.(\ref{En}) as \cite{Matloob}
\begin{eqnarray}\label{En2}
 \Delta E_{n}=\frac{ze^{4}\hbar}{24\pi^{2}\varepsilon_{0}^{2}c^{3}m^{2}}\,\mid\psi(0)\mid^{2}\int_{0}^{\frac{mc}{\hbar}}\frac{q}{q^{2}+
  \gamma^{2}}\,\sum_{\alpha}\,\mbox{Im}[\mathcal{D_{\alpha\alpha}}(\mathbf{r_{0}},\mathbf{r_{0}},\omega)]dq.
\end{eqnarray}
Now using Eqs.(\ref{IDrr},\ref{IDtt},\ref{IDzz}), we find
\begin{eqnarray}\label{Impart}
\mbox{Im}[\,\sum_{\alpha}\mathcal{D}_{\alpha\alpha}(\mathbf{r}_{0},\mathbf{r}_{0},\omega)]=\,\sum_{n=0}^{p-1}
\,2\,[\frac{\sin\,x}{x}-\,\sin^2(\frac{n\pi}{p})\,(\frac{\sin\,x}{x}-\,\frac{\cos\,x}{x^2}+\,\frac{\sin\,x}{x^3})
-\,(\frac{\sin\,x_{\theta}}{x_{\theta}}+2\,\frac{\cos\,x_{\theta}}{x_{\theta}^2}-2\,\frac{\sin\,x_{\theta}}{x_{\theta}^3})].
\end{eqnarray}
By substituting Eq.(\ref{Impart}) into Eq.(\ref{En2}), we find an expression for the energy level shifts inside the wedge as
\begin{equation}\label{21}
\frac{\Delta E_{n}}{\Delta E_{n}^{0}}=(\ln\frac{mc}{\hbar\gamma})^{-1}\,\sum_{n=0}^{p-1}\,\int_{0}^{\frac{mc}{\hbar}}\frac{q\,dq}{q^{2}+\gamma^{2}}
\,[\frac{\sin\,x}{x}-\,\sin^2(\frac{n\pi}{p})\,(\frac{\sin\,x}{x}-\,\frac{\cos\,x}{x^2}+\,\frac{\sin\,x}{x^3})
-\,(\frac{\sin\,x_{\theta}}{x_{\theta}}+2\,\frac{\cos\,x_{\theta}}{x_{\theta}^2}-2\,\frac{\sin\,x_{\theta}}{x_{\theta}^3})],
\end{equation}
which for the special choice $p=1$ tends to the familiar expression for an ideal conducting plane \cite{Matloob}.

In Fig.5, the relative energy level shifts of an atom inside the wedge are depicted in terms of the distance from the $z$-axis along the symmetry line for different apex angles $\alpha$. We see that when $\alpha=\frac{\pi}{3}$, at the region near to the narrow end of the wedge, the energy level shifts are much smaller compared to the vacuum case which means that in this region the atom is more stable in its excited state. For the more realistic case, where there are good conductors instead of ideal ones, the Green's dyadic should be obtained in the presence of matter with frequency dependent dielectric function where plasmonic effects are important and the results drastically change near the conductors \cite{casimir p}.

Using Eqs. (\ref{effective-pro}, \ref{Impart}), we can define the potential energy $\varepsilon$ for a polarizable point particle or an atom as \cite{casimir p}
\begin{equation}
\varepsilon=-\frac{1}{2}\alpha(0)<\textbf{E}^2>,
\end{equation}
where $\alpha(0)$ is the static electric dipole polarizability of the particle. In compact notation we have
\begin{equation}\label{E}
<\textbf{E}^2>=<E_{r}^2>+<E_{\theta}^2>+<E_{z}^2>.
\end{equation}
One can obtain the expression for the Casimir-Polder energy between the wedge and the atom. The force F is derivable as the gradient of the particle's potential energy. The calculation of this energy have been done in a different approach in \cite{casimir p}. However, a small check of our calculations for the Green's dyadic in the presence of a wedge shows that when $p=1$ ($\alpha=\pi$) we recover the expected Casimir-Polder result for an atom above an ideal conducting plane
\begin{equation}
\varepsilon(d)=-\frac{3}{32}\frac{\alpha(0)}{\pi^{2}d^4},
\end{equation}
where $d=r_0\sin\theta_0$ denotes the distance between the atom and the plane. This is the usual expression for the Casimir-Polder interaction \cite{Skipsey, milton c-p}.

It should be noted that for the special case $\alpha=2\pi$, we have $p=\frac{1}{2}$, that is the wedge degenerates into a half-sheet. In this case $p$ is not an integer, so we can not use Eq.(\ref{Graf}). In Sec. V, using the Green's dyadic reported in \cite{Sawaya}, we find the decay rate and energy level shifts of an atom in the vicinity of an ideal conducting half-sheet.
\section{The decay rate of an atom between parallel plates}
 There has been increasing interest in computation of the decay rate of an excited atom located between two parallel plates. At first, Barton discussed extensively the QED of charged particles between conducting plates \cite{Barton3, Barton4}. Also, Experimentally, Hulet, Hilfer, and Kleppner reported inhibited spontaneous emission by a Rydberg atom \cite{experiment}.

In the limiting case $\alpha\rightarrow0$, $r\rightarrow\infty$ such that $r\alpha=d$, the wedge geometry tends to two parallel plates separated by a distance d, so we can find the decay rates of an excited atom in this case by taking the limit of the wedge results. Using Eqs.(\ref{gammaz}, \ref{gammatheta}), the results for the parallel and perpendicular polarizations are depicted in Figs. 6 and 7 in terms of the dimensionless variables $\frac{y_0}{\lambda}$, $\frac{d}{\lambda}$, where $y=r\sin\theta_0\rightarrow r\theta_0$, denotes the distance to the lower plate, $d$ is the distance between the plates and $\lambda$ is the transition wavelength. For the case of parallel polarization, a strong suppression occurs for $\frac{d}{\lambda}<\frac{1}{2}$ since the decay rate is proportional to the density of modes in free space and in a cavity formed by two infinite conducting plates, the mode density for the electric field parallel to the surface vanishes for $\frac{d}{\lambda}<\frac{1}{2}$ \cite{two plates}.

In Figs. 6(a) and 7(a) the two curves are analogous in the sense that both are symmetric with respect to the equidistant point to the plates. However, there is a remarkable difference since the regions of enhancement of the former correspond to regions of suppression of the latter and vice versa. Also in Figs. 6(b) and 7(b), the distances between successive peaks for $\frac{\Gamma_{\parallel}}{\Gamma_{0}}$ and local minima of
$\frac{\Gamma_{\perp}}{\Gamma_{0}}$ are $\lambda$. These results are in agreement with those reported in \cite{two plates, experiment}.
\begin{figure}\label{parallel}
    \includegraphics[scale=0.4]{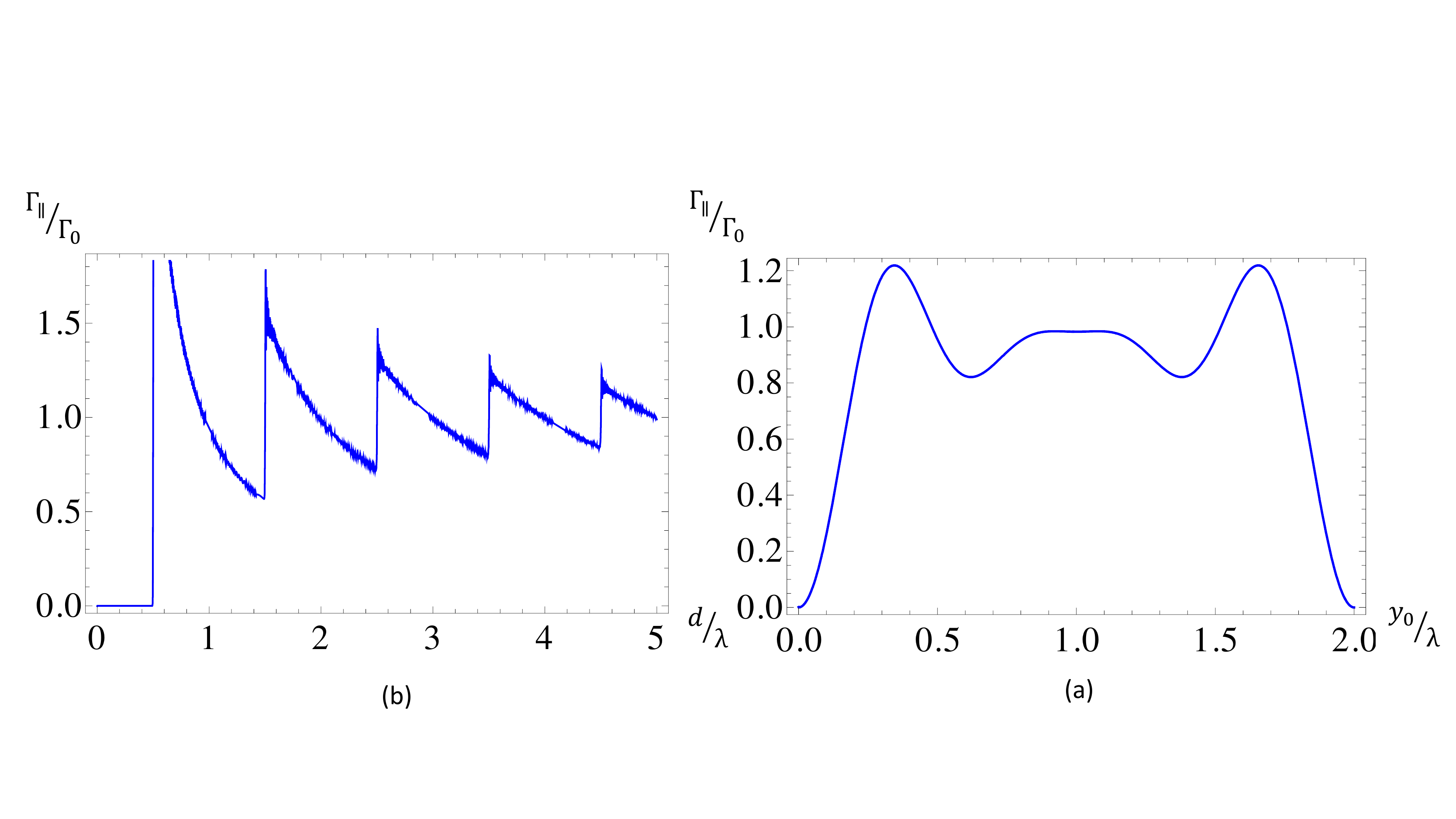}\\
  \caption{(Color online) The dimensionless decay rate for parallel polarization of the excited
atom between conducting plates in terms of the dimensionless variables $(a)\,\frac{y}{\lambda},\,(b)\,\frac{d}{\lambda}$.}\label{Fig11}
\end{figure}
\begin{figure}\label{perpendicular}
    \includegraphics[scale=0.4]{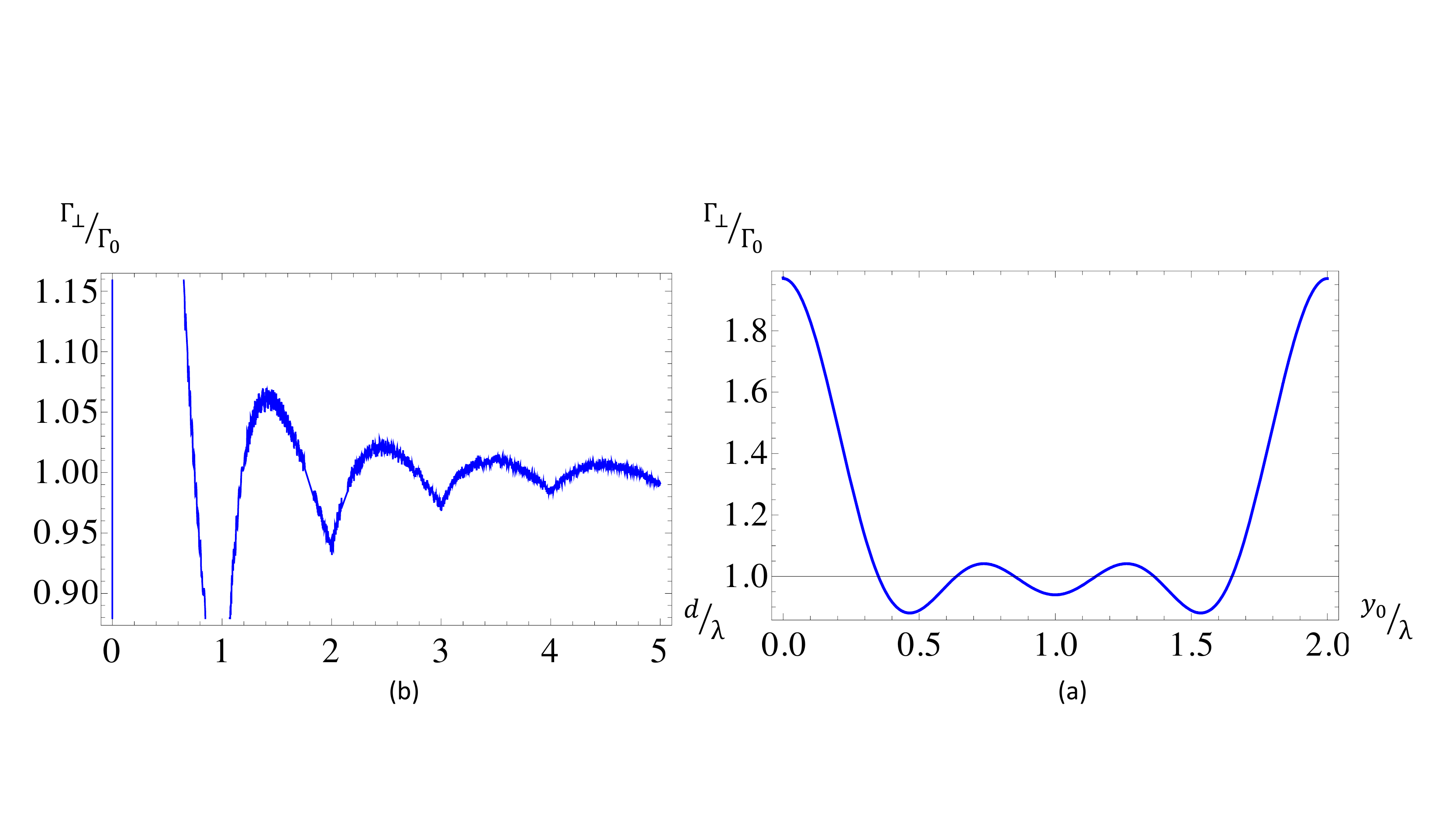}\\
  \caption{(Color online) The dimensionless decay rate for perpendicular polarization of the excited
atom between conducting plates in terms of the dimensionless variables $(a)\,\frac{y}{\lambda},\,(b)\,\frac{d}{\lambda}$.}\label{Fig10}
\end{figure}
\section{The Conducting half-sheet}
\begin{figure}
    \includegraphics[scale=0.4]{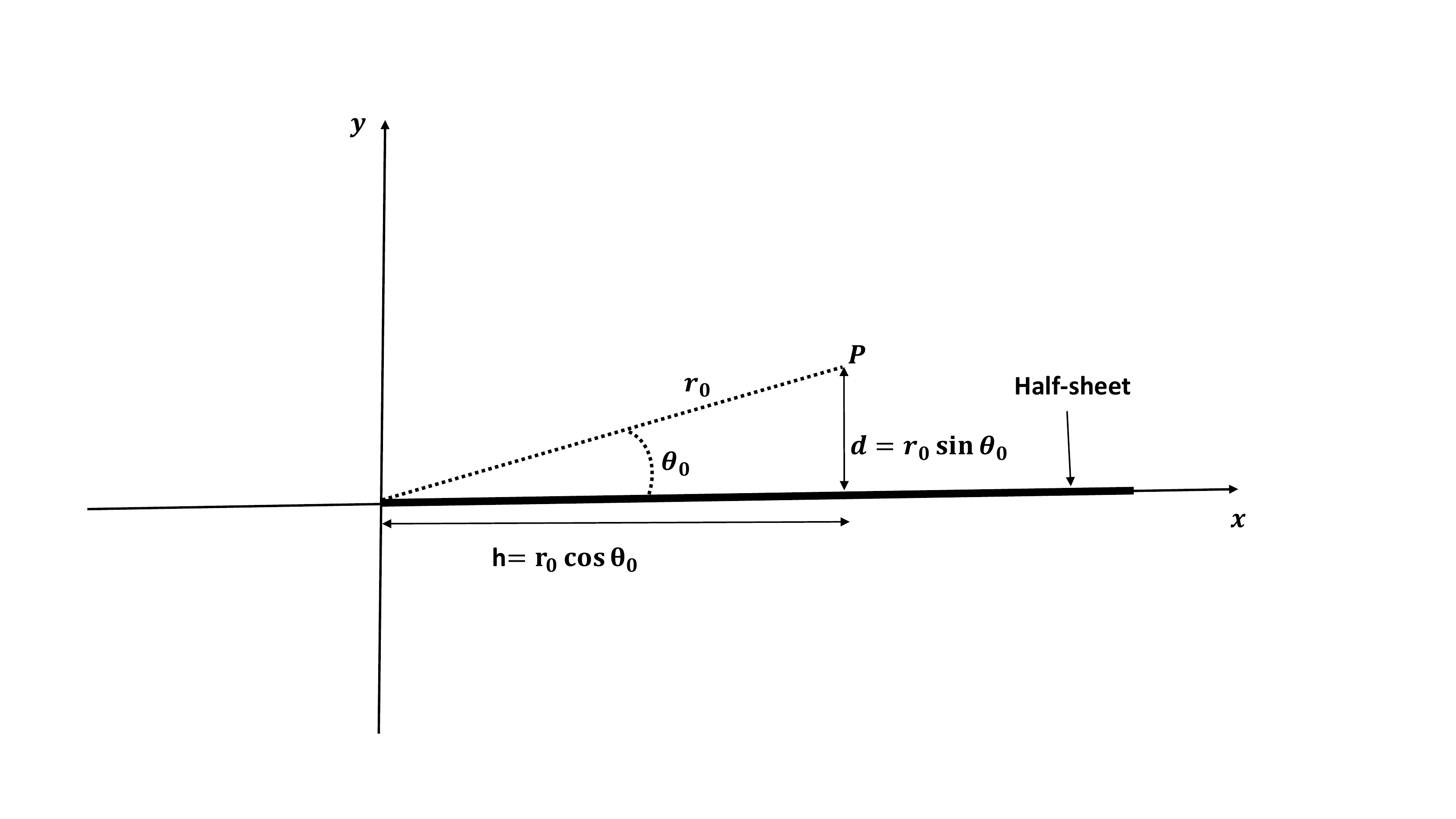}\\
  \caption{(Color online) The geometry of a conducting half-sheet.}\label{Fig7}
\end{figure}
\subsection{Green tensor}
To find the decay rate and energy level shifts of an atom near a conducting half-sheet, let us consider the geometry depicted in Fig.8. The conducting half-sheet is defined by the $xz$ plane for $x\geq0$. For this geometry, the electric type dyadic Green's function or Green's tensor $\mathcal{D}_{e1}$, satisfying boundary conditions on the walls, is given by \cite{{Sawaya,Tai1,Tai2}}
\begin{eqnarray}\label{Dyad1}
\mathcal{D}_{e1} &=&\frac{1}{2q}\,\{[(I+\frac{\nabla'\nabla'}{q^2})\,[\frac{e^{-iq r_{i}}}{r_{i}}-iq I(\zeta_{-},r_{i})]-[I_{r}.(I+\frac{\nabla'\nabla'}{q^2})][\frac{e^{-iq r_{r}}}{r_{r}}-iq I(\zeta_{+},r_{r})]\}\nonumber\\
 &-&\frac{i}{q}(\widehat{x}\,\sin(\frac{\theta}{2})-\widehat{y}\cos(\frac{\theta}{2}))
 [(\widehat{x}\,\sin(\frac{\theta'}{2})-\widehat{y}\cos(\frac{\theta'}{2}))\,\frac{H^{(2)}_{0}(q\,p)}{\sqrt{rr'}}+\frac{1}{q}
\nabla'\sin(\frac{\theta'}{2})\,\frac{r+r'}{\sqrt{rr'}}\,\frac{H^{(2)}_{1}(q\,p)}{p})],
\end{eqnarray}
where
\begin{eqnarray}
&& I = \widehat{x}\widehat{x}+\widehat{y}\widehat{y}+\widehat{z}\widehat{z}, \\
&& I_{r} = \widehat{x}\widehat{x}-\widehat{y}\widehat{y}+\widehat{z}\widehat{z}, \\
&& p =\sqrt{(r+r')^2+(z-z')^2},\\
&& r_{i} =\sqrt{(x-x')^2+(y-y')^2+(z-z')^2}=\sqrt{r^2+r'^2-2\,rr'\,\cos(\theta-\theta')+(z-z')^2},\\
&& r_{r} =\sqrt{(x-x')^2+(y+y')^2+(z-z')^2}=\sqrt{r^2+r'^2-2\,rr'\,\cos(\theta+\theta')+(z-z')^2},\\
&& I(\zeta,\eta)= \int_{0}^{\zeta}\,dt\,\frac{H^{(2)}_{1}(\omega\,\sqrt{t^2+\eta^2})}{\sqrt{t^2+\eta^2}},
\end{eqnarray}\label{20}
and $\zeta_{\mp}=2\,\sqrt{rr'}\,\cos(\frac{\theta\mp\theta'}{2})$.

By making use of Eq.(\ref{Dyad1}), we find the diagonal components of Green's tensor as
\begin{eqnarray}\label{55}
 \mathcal{D}_{yy}&=&\frac{1}{2q}\,[(1+\frac{1}{q^2}\frac{\partial}{\partial y'}\frac{\partial}{\partial y'})(\frac{e^{-iq r_{i}}}{r_{i}}-iq I(\zeta_{-},r_{i}))+(1+\frac{1}{q^2}\frac{\partial}{\partial y'}\frac{\partial}{\partial y'})(\frac{e^{-iq r_{r}}}{r_{r}}-iq I(\zeta_{+},r_{r}))]\nonumber\\
 &-&\frac{i}{q}[\cos(\frac{\theta}{2})\,\cos(\frac{\theta'}{2})\,\frac{J_{0}(q\,p)}{\sqrt{rr'}}-\frac{1}{q}
 \,\cos(\frac{\theta}{2})\,\frac{\partial}{\partial y'}(\sin(\frac{\theta'}{2})\,\frac{r+r'}{\sqrt{rr'}}\,\frac{J_{1}(q\,p)}{p})],
\end{eqnarray}
\begin{eqnarray}\label{56}
 \mathcal{D}_{xx}&=&\frac{1}{2q}\,[(1+\frac{1}{q^2}\frac{\partial}{\partial x'}\frac{\partial}{\partial x'})(\frac{e^{-iq r_{i}}}{r_{i}}-iq I(\zeta_{-},r_{i}))-(1+\frac{1}{q^2}\frac{\partial}{\partial x'}\frac{\partial}{\partial x'})(\frac{e^{-iq r_{r}}}{r_{r}}-iq I(\zeta_{+},r_{r}))]\nonumber\\
 &-&\frac{i}{q}[\sin(\frac{\theta}{2})\,\sin(\frac{\theta'}{2})\,\frac{J_{0}(q\,p)}{\sqrt{rr'}}+\frac{1}{q}
 \,\sin(\frac{\theta}{2})\,\frac{\partial}{\partial x'}(\sin(\frac{\theta'}{2})\,\frac{r+r'}{\sqrt{rr'}}\,\frac{J_{1}(q\,p)}{p})],
\end{eqnarray}
\begin{equation}\label{57}
 \mathcal{D}_{zz} =\frac{1}{2q}\,\{(1+\frac{1}{q^2}\frac{\partial}{\partial z'}\frac{\partial}{\partial z'})(\frac{e^{-iq r_{i}}}{r_{i}}-iq I(\zeta_{-},r_{i}))-(1+\frac{1}{q^2}\frac{\partial}{\partial z'}\frac{\partial}{\partial z'})(\frac{e^{-iq r_{r}}}{r_{r}}-iq I(\zeta_{+},r_{r}))\}.
\end{equation}
\subsection{The decay rate}
\begin{figure}
    \includegraphics[scale=0.4]{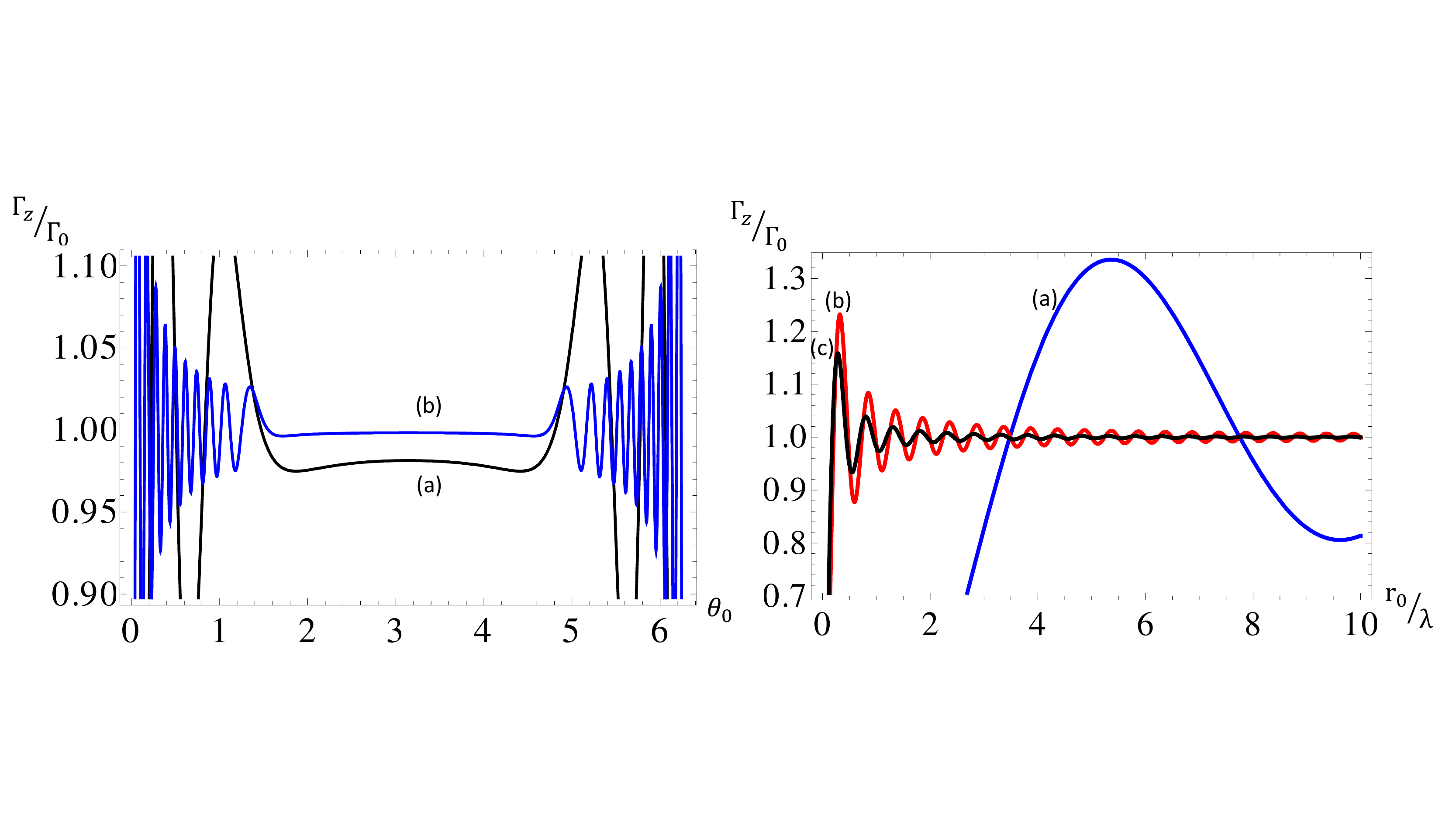}\\
 \caption{(Color online) The dimensionless decay rate $\frac{\Gamma_{z}}{\Gamma_{0}}$ of an excited atom in the vicinity of a half-sheet (Left Figure) in terms of the angle $\theta_0\in(0, 2\pi)$, for fixed distances (a) $\frac{r_{0}}{\lambda}=1 $ and (b) $\frac{r_{0}}{\lambda}=5$, (Right Figure) along the lines (a) $\theta_0=\frac{\pi}{50}$, (b) $\theta_0=\frac{\pi}{2}$ and (c) $\theta_0=\frac{5\pi}{6}$.}\label{Fig12}
\end{figure}
\begin{figure}
    \includegraphics[scale=0.4]{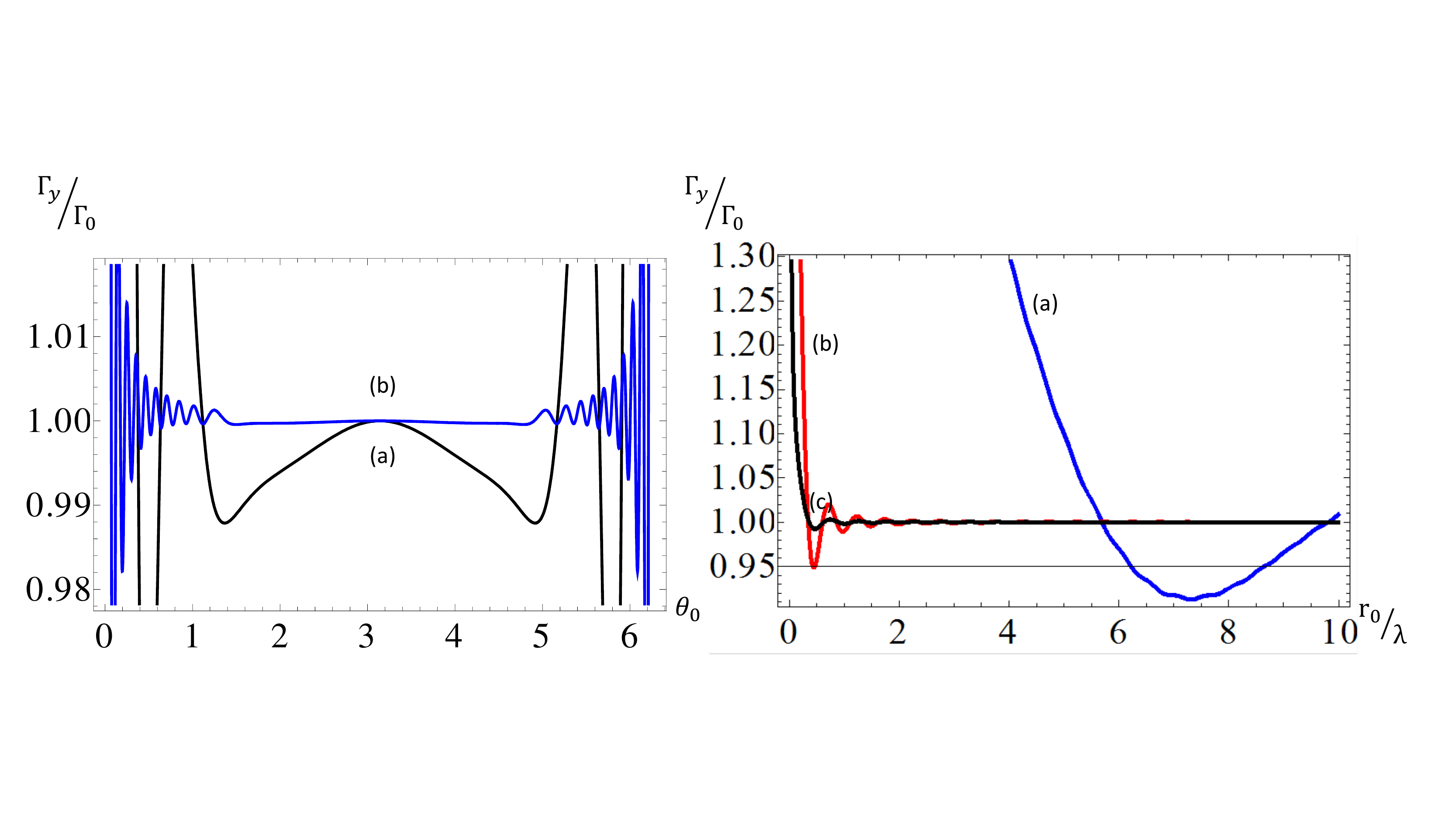}\\
  \caption{(Color online) The dimensionless decay rate $\frac{\Gamma_{y}}{\Gamma_{0}}$ of an excited atom in the vicinity of a half-sheet (Left Figure) in terms of the angle $\theta\in(0, 2\pi)$, for fixed distances (a) $\frac{r_{0}}{\lambda}=1 $ and (b) $\frac{r_{0}}{\lambda}=5$, (Right Figure) along the lines (a) $\theta_0=\frac{\pi}{50}$, (b) $\theta_0=\frac{\pi}{2}$ and (c) $\theta_0=\frac{5\pi}{6}$.}\label{Fig13}
\end{figure}
To calculate $\Gamma_{\perp}$($\Gamma_{y}$ in our notation), we need the $yy$-component of the Green's tensor, by inserting Eq.(\ref{55}) into Eq.(\ref{Gama}), we find
\begin{eqnarray}\label{gy}
\frac{\Gamma_{y}}{\Gamma_{0}}=&-&\frac{3}{4q}\,\mbox{Im}\bigg\{[(1+\frac{1}{q^2}\frac{\partial}{\partial y'}\frac{\partial}{\partial y'})(\frac{e^{-iq r_{r}}}{r_{r}}-iq I(\zeta_{+},r_{r}))+(1+\frac{1}{q^2}\frac{\partial}{\partial y'}\frac{\partial}{\partial y'})(\frac{e^{-iq r_{i}}}{r_{i}}-iq I(\zeta_{-},r_{i}))]\nonumber\\
 &-& 2i\,[\cos(\frac{\theta}{2})\,\cos(\frac{\theta'}{2})\,\frac{J_{0}(q\,p)}{\sqrt{rr'}}-\frac{1}{q}
 \,\cos(\frac{\theta}{2})\,\frac{\partial}{\partial y'}(\sin(\frac{\theta'}{2})\,\frac{r+r'}{\sqrt{rr'}}\,\frac{J_{1}(q\,p)}{p})]\bigg\},\bigg|_{x\rightarrow x',y\rightarrow y',z\rightarrow z'}
\end{eqnarray}
and for the in-plane polarizations $\Gamma_{x}$ or $\Gamma_{z}$ we will find
\begin{eqnarray}\label{gx}
\frac{\Gamma_{x}}{\Gamma_{0}}&=&\frac{3}{4q}\,\mbox{Im}\bigg\{[(1+\frac{1}{q^2}\frac{\partial}{\partial x'}\frac{\partial}{\partial x'})(\frac{e^{-iq r_{r}}}{r_{r}}-iq I(\zeta_{+},r_{r}))-(1+\frac{1}{q^2}\frac{\partial}{\partial x'}\frac{\partial}{\partial x'})(\frac{e^{-iq r_{i}}}{r_{i}}-iq I(\zeta_{-},r_{i}))]\nonumber\\
 &+& 2i\,[\sin(\frac{\theta}{2})\,\sin(\frac{\theta'}{2})\,\frac{J_{0}(q\,p)}{\sqrt{rr'}}+\frac{1}{q}
 \,\sin(\frac{\theta}{2})\,\frac{\partial}{\partial x'}(\sin(\frac{\theta'}{2})\,\frac{r+r'}{\sqrt{rr'}}\,\frac{J_{1}(q\,p)}{p})]\bigg\}\bigg|_{x\rightarrow x',y\rightarrow y',z\rightarrow z'},
\end{eqnarray}
\begin{equation}\label{gz}
\frac{\Gamma_{z}}{\Gamma_{0}}=\frac{3}{4q}\,\mbox{Im}\bigg[(1+\frac{1}{q^2}\frac{\partial}{\partial z'}\frac{\partial}{\partial z'})(\frac{e^{-iq r_{r}}}{r_{r}}-iq I(\zeta_{+},r_{r}))-(1+\frac{1}{q^2}\frac{\partial}{\partial z'}\frac{\partial}{\partial z'})(\frac{e^{-iq r_{i}}}{r_{i}}-iq I(\zeta_{-},r_{i}))\bigg]\bigg|_{x\rightarrow x',y\rightarrow y',z\rightarrow z'}.
\end{equation}
Using Eqs. (\ref{gy},\ref{gz}) and doing straightforward calculations, we find
\begin{eqnarray}\label{gamay}
\frac{\Gamma_{y}}{\Gamma_{0}}&=&\frac{1}{2}-\frac{3}{2}\,\bigg[\frac{\cos\,(2\,y_{0}q)}{(2\,y_{0}q)^2}-\frac{\sin (2\,y_{0}q)}{(2\,y_{0}q)^3}
+(2\,\cos(\theta_{0}/2)\,\cos(3\theta_{0}/2)-\frac{\cos(\theta_{0})}{2}-\frac{\cos(2\theta_{0})}{2}+\frac{\sin^2(\theta_{0})}{2})\,
\frac{J_{1}(2\,r_{0}q)}{(2\,r_{0}q)^2}\nonumber\\
&& -(\frac{\sin^2(\theta_{0})}{4})\,\frac{J_{2}(2\,r_{0}q)}{(2\,r_{0}q)}-(2\cos^4(\theta_{0}/2)+\frac{\sin^2(\theta_{0})}{4})\,
\frac{J_{0}(2\,r_{0}q)}{(2\,r_{0}q)}-\frac{1}{2}\,\int_{0}^{2\,r_{0}q}\,dx\,
(\frac{J_{1}(x)}{x}-\frac{J_{2}(x)}{x^2})\nonumber\\
&& -\frac{1}{2}\int_{0}^{2\,r_{0}q\,cos(\theta_{0})}\,dx\,\frac{[z^3J_{0}(z)-(2z^2+(2\,y_{0}q)^2z^2-z^4)J_{1}(z)
+4(2\,y_{0}q)^2zJ_{2}(z)]}{z^5}
\bigg]
\end{eqnarray}
\begin{eqnarray}\label{gamaz}
\frac{\Gamma_{z}}{\Gamma_{0}}&=&\frac{1}{2}-\frac{3}{4}\,\bigg[\frac{\sin (2\,y_{0}q)}{(2\,y_{0}q)}+\frac{\cos\,(2\,y_{0}q)}{(2\,y_{0}q)^2}-\frac{\sin (2\,y_{0}q)}{(2\,y_{0}q)^3}+\,\int_{0}^{2\,r_{0}q\,cos(\theta_{0})}\,dx\,[\frac{J_{1}(z)}{z}-
\frac{J_{1}(z)}{z^3}\nonumber\\
 && +\frac{J_{0}(z)}{2z^2}-
\frac{J_{2}(z)}{2z^2}]-\int_{0}^{2\,r_{0}q}\,dx\,[\frac{J_{1}(x)}{x}-
\frac{J_{1}(x)}{x^3}+\frac{J_{0}(x)}{2\,x^2}-
\frac{J_{2}(x)}{2\,x^2}]\bigg]
\end{eqnarray}
where $ y_{0}=r_{0}\sin\theta_{0}$ and $ z=\sqrt{x^2+(2\,y_{0}q)^2}$.

The decay rate for these two polarizations in terms of the dimensionless distance $\frac{r_{0}}{\lambda}$ along the lines (a) $\theta_0=\frac{\pi}{50}$, (b) $\theta_0=\frac{\pi}{2}$ and (c) $\theta_0=\frac{5\pi}{6}$ is depicted in Figs. 9(Right figure) and 10(Right figure). The decay rates in terms of $\theta$ for fixed distances is also depicted in Figs. 9(Left figure) and 10(Left figure), showing a symmetrical behaviour around $\theta_0=\pi$, as expected. It is interesting to note that in these figures, the presence of the half-sheet affects the decay rate only for angles $\theta_0<\pi/2$ or $\theta_0>3\pi/2$, this is because the emitted photon from the excited atom inside the angle $\pi/2<\theta_0<3\pi/2$ can not be reflected back to the atom.

Let us as a consistency check, find the limiting case where in Fig. 8, the atom is placed at a distance far from the $z$-axis or the edge of the half-sheet, that is $d$ is fixed and $ h\rightarrow\infty$. In this limiting case using Eq.(\ref{gradshteyn}) we find
\begin{equation}
\frac{\Gamma_{y}}{\Gamma_{0}}=-\frac{3}{4q}\,[(1+\frac{1}{q^2}\frac{\partial}{\partial y'}\frac{\partial}{\partial y'})(\frac{-2\sin(q r_{r})}{r_{r}})+(1+\frac{1}{q^2}\frac{\partial}{\partial y'}\frac{\partial}{\partial y'})(\frac{-2\sin(q r_{i})}{r_{i}})]|_{x\rightarrow x',y\rightarrow y',z\rightarrow z'},
\end{equation}
\begin{equation}
\frac{\Gamma_{x}}{\Gamma_{0}}=\frac{3}{4q}\,[(1+\frac{1}{q^2}\frac{\partial}{\partial x'}\frac{\partial}{\partial x'})(\frac{-2\sin(q r_{r})}{r_{r}})-(1+\frac{1}{q^2}\frac{\partial}{\partial x'}\frac{\partial}{\partial x'})(\frac{-2\sin(q r_{i})}{r_{i}})]|_{x\rightarrow x',y\rightarrow y',z\rightarrow z'},
\end{equation}
\begin{equation}
\frac{\Gamma_{z}}{\Gamma_{0}}=\frac{3}{4q}\,[(1+\frac{1}{q^2}\frac{\partial}{\partial z'}\frac{\partial}{\partial z'})(\frac{-2\sin(q r_{r})}{r_{r}})-(1+\frac{1}{q^2}\frac{\partial}{\partial z'}\frac{\partial}{\partial z'})(\frac{-2\sin(q r_{i})}{r_{i}})]|_{x\rightarrow x',y\rightarrow y',z\rightarrow z'}.
\end{equation}
by taking the derivatives and evaluating the expressions at $\mathbf{r}'=\mathbf{r}$, we finally find
\begin{equation}
\frac{\Gamma_{y}}{\Gamma_{0}}=1-3[\frac{\cos(2q d)}{(2q d)^2}-\frac{\sin(2q d)}{(2q d)^3}],
\end{equation}
and
\begin{equation}
\frac{\Gamma_{x}}{\Gamma_{0}}=\frac{\Gamma_{z}}{\Gamma_{0}}=1-\frac{3}{2}[\frac{\sin(2q d)}{(2q d)}+\frac{\cos(2q d)}{(2q d)^2}-\frac{\sin(2q d)}{(2q d)^3}],
\end{equation}
which are the same results reported in \cite{Matloob} for an ideal conducting plane, as expected.

As another consistency check let us find the Casimir force on the atom when it is polarized perpendicular to the half-sheet and directly above the edge. When atom is polarizable only in the $y$ direction (see Fig. 8), the only component of Green's dyadic that contributes is $yy$-component. We insert this component into Eq.(\ref{E}) and the result of a straightforward calculation  at $\theta=\pi$ leads to
\begin{equation}
<\textbf{E}^2>=<E_{y}^2>=0.
\end{equation}
This means that when the atom is polarized perpendicular to the half-sheet and located exactly above the edge of half-sheet, there is no force on the atom, in agreement with the result reported in \cite{Eberlein}.
\subsection{The energy level shifts}
By inserting Eqs.(\ref{55},\ref{56},\ref{57}) into Eq.(\ref{En}), we can find the energy level shift of an atom in the presence of an ideal conducting half-sheet as
\begin{equation}\label{21}
\frac{\Delta E_{n}}{\Delta E_{n}^{0}}=\frac{1}{4\ln\frac{mc}{\hbar\gamma}}\int_{0}^{\frac{mc}{\hbar}}\frac{dq}{q^{2}+\gamma^{2}}\,A,
\end{equation}
where
\begin{eqnarray}\label{A}
A &=& \mbox{Im}\bigg\{[1+\frac{1}{q^2}(\frac{\partial}{\partial x'}\frac{\partial}{\partial x'}-\frac{\partial}{\partial y'}\frac{\partial}{\partial y'}
+\frac{\partial}{\partial z'}\frac{\partial}{\partial z'})][\frac{e^{-iq r_{r}}}{r_{r}}-iq I(\zeta_{+},r_{r})]\nonumber\\
&&-[3+\frac{1}{q^2}(\frac{\partial}{\partial x'}\frac{\partial}{\partial x'}+\frac{\partial}{\partial y'}\frac{\partial}{\partial y'}+\frac{\partial}{\partial z'}\frac{\partial}{\partial z'})][\frac{e^{-iq r_{i}}}{r_{i}}-iq I(\zeta_{-},r_{i})]\nonumber\\
&&+ 2i\,[\cos(\frac{\theta-\theta'}{2})\,\frac{J_{0}(q\,p)}{\sqrt{rr'}}+\frac{1}{q}
 \,\sin(\frac{\theta}{2})\,\frac{\partial}{\partial x'}(\sin(\frac{\theta'}{2})\,\frac{r+r'}{\sqrt{rr'}}\,\frac{J_{1}(q\,p)}{p})\nonumber\\
&&-\frac{1}{q}
 \,\cos(\frac{\theta}{2})\,\frac{\partial}{\partial y'}(\sin(\frac{\theta'}{2})\,\frac{r+r'}{\sqrt{rr'}}\,\frac{J_{1}(q\,p)}{p})]\bigg\}
\bigg|_{x\rightarrow x',y\rightarrow y',z\rightarrow z'}
\end{eqnarray}
The above expression for the energy level shift cannot be reduced any further to a simple analytical form and one should invoke numerical calculations.
As before, when $d$ is fixed and $ h\rightarrow\infty$ (see Fig.86), the three last terms in Eq.(\ref{A}) tend to zero and we find
\begin{equation}\label{21}
\frac{\Delta E_{n}}{\Delta E_{n}^{0}}=1-(\ln\frac{mc}{\hbar\gamma})^{-1}\,\int_{0}^{\frac{mc}{\hbar}}\frac{q\,dq}{q^{2}+\gamma^{2}}\,[\frac{\sin(2q d)}{(2q d)}+2\frac{\cos(2q d)}{(2q d)^2}-2\frac{\sin(2q d)}{(2q d)^3}],
\end{equation}
again in agreement with the result reported in \cite{Matloob}, as expected.
\section{Conclusions}
Explicit expressions for the decay rate and energy level shifts of an atom in the presence of an ideal conducting wedge, two conducting parallel plates and a half-sheet are obtained in the frame work of the canonical quantization approach. The angular and radial dependence of the decay rate for different atomic polarizations of an excited atom and also of the energy level shifts are depicted and discussed. The consistency of the present approach in some limiting cases is investigated by comparing the relevant results obtained here to the previously reported results. For distances from the cusp smaller than a certain value determined by the opening angle of the wedge, there are configurations for which the atom will not decay at all. This is more clearly understood for the case of conducting parallel plates where for the case of a transition dipole moment parallel to the plates, a strong suppression occurs for $\frac{d}{\lambda}<\frac{1}{2}$ since the mode density for the electric field parallel to the surface vanishes for $\frac{d}{\lambda}<\frac{1}{2}$. The appearance of enhancement or inhibition of the emission depends not only on the dipole location of the atom  but also on the dipole orientational of the atom and the opening angle of the wedge. These studies open the way to devising cavity quantum electrodynamics systems, such as the storage, processing and retrieval of quantum bits for practical realization of quantum information processing \cite{qinformation}.

\end{document}